\documentstyle[prl,aps,twocolumn,epsf]{revtex}
\draft
\begin{document}

\title{Local Spectral Density for a Periodically Driven System of Coupled Quantum
States with Strong Imperfection in Unperturbed Energies}

\author{V. S. Starovoitov}

\address{B.I.Stepanov Institute of Physics, NASB, 220072, Scarina ave. 70, Minsk,\\
Belarus }

\date{\today}

\maketitle
\begin{abstract}
A random matrix theory approach is applied in order to analyze the
localization properties of local spectral density for a generic system of
coupled quantum states with strong static imperfection in the unperturbed
energy levels. The system is excited by an external periodic field, the
temporal profile of which is close to monochromatic one. The shape of local
spectral density is shown to be well described by the contour obtained from
a relevant model of periodically driven two-states system with irreversible
losses to an external thermal bath. The shape width and the inverse
participation ratio are determined as functions both of the Rabi frequency
and of parameters specifying the localization effect for our system in the
absence of external field.
\end{abstract}

\narrowtext

\section{Introduction}

The statistical properties of many-body quantum objects attract considerable
attention in a broad range of modern physics including condensed matter
physics and quantum optics. Of a special interest are the properties
specifying the temporal evolution for mesoscopic systems of quantum states
coupled by interaction. Extensive investigations of many-body interacting
systems (such as nuclei, many-electron atoms, quantum dots, quantum spin
glasses and quantum computers) have shown that the state-state interaction
can be responsible for the quantum localization effect and the quantum chaos
in isolated complex systems \cite
{ref1,ref2,ref3,ref4,ref5,ref6,ref7,ref8,ref9}. As in Random Matrix Theory
(RMT) \cite{ref10,ref11}, the level spacing statistics is well described by
the Wigner-Dyson distribution and the individual basis states are spread
over the large number of eigenstates. In a sense the interaction leads to
dynamic thermalization without coupling to an external thermal bath. On the
other hand, to our knowledge, there are no direct theoretical studies for
mesoscopic systems driven by an external field.

In the work we apply a RMT approach in order to investigate the localization
effect for a complex system of coupled quantum states in the presence of an
external field. In the approach a single matrix from a given statistical
ensemble (characterized by a few matrix parameters) represents the
Hamiltonian ${\bf H}\left( t\right) $ for the typical (generic) dynamical
system of given class. Our system is represented by a matrix, which includes
a leading diagonal with disordered random values and random off-diagonal
elements inside a band with the halfwidth $b$. This band random matrix with
disordered diagonal (BRMDD) is a generic Hamiltonian model for the systems
with strong static imperfection in the unperturbed energy levels. The
imperfection plays an essential role for the quibit composition and can
destroy the operability of quantum computer \cite{ref9,ref9A,ref9B}.
BRMDD-based models were applied to study the electron transport problem \cite
{ElectrTransport82,et88,ElectrTransport95} and the problem of interacting
particles in a random potential \cite{BRMPB94,BRMPB95,Weinmann96}. The
results obtained with the help of these models are of an obvious interest
also in analyzing such few-freedoms physical objects as the vibrational
quasicontinuum of polyatomic molecule \cite{ref12,ref13,Wolynes}.

The statistical properties of conservative BRMDD-based systems were studied
in some detail \cite{ref14,ref15,ref16,ref17,ref18,ref19,ref20}. The shape,
the localization length and the inverse participation ratio (IPR) for
eigenfunction have been investigated with the help of numerical simulations 
\cite{ref14,ref18,ref19} and the supersymmetry approach \cite{ref19,ref20}.
These investigations have exhibited the Lorentzian shape for the local
spectral density (LSD) of states under the circumstances when a
non-perturbative localization regime is realized. The state-state
interaction strength, at which the eigenstates are extended over the whole
matrix size $N$ and the eigenenergy level spacing statistics has the
Wigner-Dyson form, have been revealed for the BRMDD with essentially large
band (when $2b+1\gg \sqrt{N}$) \cite{ref17,ref18,ref19,ref20}. The
association between ergodic properties of LSD and the matrix parameters have
been clarified recently for the BRMDD with arbitrary small band \cite{ref21}%
. The lack of ergodicity for LSD is shown in \cite{ref21} to be identified
with an exponential increase in IPR with the strength of state-state
interaction.

We study the LSD of states for a generic BRMDD-based quantum system to be
excited by an external quasi-monochromatic field. The LSD was introduced in
1955 by Wigner \cite{ref22,ref23} and successfully employed in RMT to
describe statistically the localization effects for complex quantum systems 
\cite{ref10,ref11} (including the systems represented by band random
matrices with reordered leading diagonal \cite{ref24,ref25}). In our study
this quantity specifies spreading of the energy concentrated initially in an
individual basis state $\left| g\right\rangle $, between the quasienergies
and gives the quasienergy spectrum of $\left| C\left( \omega \right) \right|
^2$ where $C\left( \omega \right) $ is the Fourier transform for correlation 
\begin{equation}
\label{Ct}C\left( t\right) =\left\langle g\right| \exp \left( -i{\bf H}%
\left( t\right) t/\hbar \right) \left| g\right\rangle . 
\end{equation}
As for the conservative systems \cite{ref10,ref11}, the shape of LSD may be
characterized by the width $\Gamma $ of a quasienergy scale, on which the
individual state $\left| g\right\rangle $ is localized. The number of
quasienergies populating this scale is given by the ratio $\Gamma /\Delta
_\omega =L$ designated here as a localization length of LSD ($\Delta _\omega 
$ is the mean quasienergy spacing). Hence the quantity $L$ specifies the
greatest possible number of quasienergies, where the basis state $\left|
g\right\rangle $ can be effectively admixed. The ergodic properties for the
system can be identified with the structure of LSD. The non-ergodic LSD is a
strongly fluctuating spiked function and the IPR $\xi $, which gives the
actual number of quasienergies involving the state $\left| g\right\rangle $,
is low in comparison with $L$. In the ergodicity case the LSD is monotonic
and the number $\xi $ approaches the value of $L$.

Here we study the shape and the ergodic features of LSD and estimate the
shape width $\Gamma $ and the IPR $\xi $ as functions of field-system
interaction strength. The functions are analyzed in relation to relevant
properties (the mean quasienergy spacing, the width, the localization length
and the IPR) specifying the localization effect for our BRMDD-based system
in the absence of external field.

\section{Model desription}

\begin{figure} [h!]
\epsfxsize= 8. cm
\centerline{\epsfbox{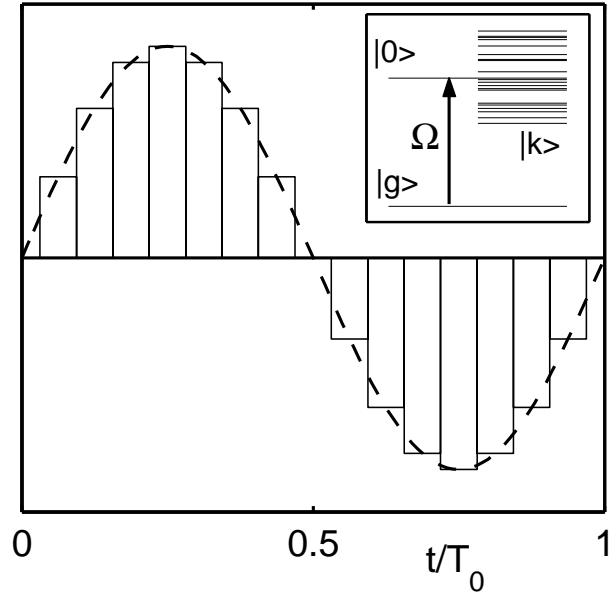}}
\vspace{3 mm}
\caption{
A temporal shape of external periodical field (solid lines, $M=16$). 
The fat dashed line gives the envelope $\sin (\omega _{f}t)$. The inset shows
a diagram of unperturbed energy levels 
for represented BRMDD-based quantum system.
}
\label{Fig1}
\end{figure}

We consider a generic system of quantum states, the unperturbed energy
levels of those are depicted in the inset of figure \ref{Fig1}. The system
consists of a single state $\left| g\right\rangle $ and $N=2K+1$ states $%
\left| k\right\rangle $ ($k=-K,\ldots ,K$). The total Hamiltonian of the
system is considered as the sum

\begin{equation}
\label{Htot}{\bf H}\left( t\right) ={\bf H}^{(0)}+{\bf H}^{\left( \Omega
\right) }\left( t\right) . 
\end{equation}
Here the time-independent operator ${\bf H}^{(0)}$ describes a state-state
interaction in the absense of field. The time-dependent part ${\bf H}%
^{\left( \Omega \right) }\left( t\right) $ specifies a field-system
interaction.

In the basis of unperturbed states the Hamiltonian ${\bf H}^{(0)}$ is
represented in terms of a real symmetric matrix with statistically
independent random elements 
\begin{equation}
\label{H0}\left\langle i\right| {\bf H}^{(0)}\left| j\right\rangle
=E_i^{(0)}\delta_{ij} +\left\langle i\right| {\bf V}\left|
j\right\rangle 
\end{equation}
where the off-diagonal elements $\left\langle i\right| {\bf V}\left|
j\right\rangle =V_{ij}=V_{ji}$ specify the state-state interaction 
($ i,j=g,-K,\ldots ,K$). In the
model we take into account an interaction between the states $\left|
k\right\rangle $. The corresponding elements $V_{k^{\prime }k}$ are
distributed uniformly in the interval $[-V,V]$ with $\left\langle
V_{k^{\prime }k}\right\rangle =0$ and $\left\langle V_{k^{\prime
}k}^2\right\rangle =V^2/3=v^2$ if $\left| k^{\prime }-k\right| <b$ or are
zero otherwise. The coupling of the state $\left| g\right\rangle $ with $%
\left| k\right\rangle $ is ignored and $V_{gk}=V_{kg}=0$. The diagonal
elements $E_k^{(0)}$ corresponding to the energy levels of states $\left|
k\right\rangle $ are uniformly distributed according to the Poisson
statistics with the mean spacing $\Delta $ between adjacent levels: $%
-K\Delta \leq E_k^{(0)}\leq K\Delta $. One of the energy levels (for
definiteness, we take the level $E_0^{(0)}$ for the state $\left|
0\right\rangle $) is located in the midpoint of interval $\left[ -K\Delta
,-K\Delta \right] $: $\left\langle E_k^{(0)}\right\rangle \approx E_0^{(0)}=0
$. The state $\left| g\right\rangle $ is the lowest one. The energy levels
are chosen so that $E_0^{(0)}-E_g^{(0)}\gg 2K\Delta $. Note that in the
basis of states $\left| k\right\rangle $ the operator ${\bf H}^{(0)}$ is
represented by a BRMDD.

\vspace{3 mm}

An external periodic field with the frequency $\omega _f=E_0^{(0)}-E_g^{(0)}$
stimulates the transitions between the non-perturbed basis states $\left|
g\right\rangle $ and $\left| 0\right\rangle $. The field-induced coupling
between $\left| g\right\rangle $ and the states $\left| k\right\rangle $
with $k\neq 0$ is accepted to be negligible. The temporal profile of the
field is shown in figure 1 to be a piecewise function with the envelope $%
\sin (\omega _ft)$. The total number of pieces covering the field period
interval $T_f=2\pi /\omega _f$ is equal to $M$. At large $M$ the field shape
is close to monochromatic one. On each of the time pieces $\left[
t_m,t_{m+1}\right] $ (the integer $m$ gives the number of piece interval)
the field is time-invariant and the field-system interaction is represented
by 
\begin{equation}
\label{Hom}{\bf H}^{\left( \Omega \right) }\left( t_m\right) =\Omega \sin
(2\pi m/M){\bf D.} 
\end{equation}
Here the Rabi frequency $\Omega $ specifies the strength of field-system
interaction. The operator ${\bf D}$ specifies the field-induced coupling
between the states: 
\begin{equation}
\label{D}\left\langle i\right| {\bf D}\left| j\right\rangle =
\delta_{i0}\delta_{gj}+\delta_{ig}\delta_{0j}. 
\end{equation}
At such an approximation the evolution of system for one field period
interval is described by a unitary operator 
\begin{equation}
\label{U}{\bf U}(T_f)=\prod_{m=1}^M\exp \left( -i\left[ {\bf H}\left(
t_m\right) \right] T_f/M\hbar \right) . 
\end{equation}
From diagonalization of the matrix ${\bf U}(T_f)$ we obtain the
eigenfunctions $\left| \alpha _j\right\rangle $ and the quasienergies $%
\omega _j$. The quantity $W_{jg}=|\left\langle \alpha _j|g\right\rangle |^2$
gives a probability to find the probing basis state $\left| g\right\rangle $
in the eigenstate $\left| \alpha _j\right\rangle $.

The described model can be considered as a driven two-state system (the
states $\left| g\right\rangle $ and $\left| 0\right\rangle $), the upper
state $\left| 0\right\rangle $ of which is coupled by interaction with a
number of others quantum states. The interaction-induced probability
distribution of the state $\left| 0\right\rangle $ over the eigenstates of $%
{\bf H}^{(0)}$ corresponds to a quasienergetic distribution of $\left|
0\right\rangle $ in the absence of field ($\Omega =0$). Since the operator $%
{\bf H}^{(0)}$ is given in terms of a BRMDD the probability distribution of $%
\left| 0\right\rangle $ over quasienergy scale is described by the
Breight-Wigner shape (that is, by the Lorentzian contour) \cite
{ref14,ref18,ref19,ref21} 
\begin{equation}
\label{eqLor}\rho _0\left( \omega \right) =\frac{\Gamma _0}{2\pi }\frac
1{\omega ^2+\Gamma _0^2/4}
\end{equation}
of the width $\Gamma _0$. In some sense the state $\left| 0\right\rangle $
is Lorentzian-broadened. Therefore we associate our BRMDD-based system with
a two-state system, the upper state of which is continuously
Lorentzian-broadened with the width $\Gamma _0$ because of irreversible
losses to an external thermal bath. Here the lower state has no losses and
hence is not broadened. This two-state system with irreversible losses
(TSSIL) is stimulated by an external monochromatic field. Such a well-known
model (see, for instance, \cite{ref26,ref27}) corresponds to our BRMDD-based
system in the limit of infinitely large number $N$ of coupled states $\left|
k\right\rangle $.

Depending on the ratio between the width $\Gamma _0$ and the Rabi frequency $%
\Omega $ the TSSIL model gives two shapes of $\left| C\left( \omega \right)
\right| ^2$ \cite{ref26,ref27}. At $2\Omega /\Gamma _0<1$ the correlation $%
C(t)$ is an decaying aperiodical function of time. Then the shape of $\left|
C\left( \omega \right) \right| ^2$ is described by a contour 
\begin{equation}
\label{eq1}\left| C\left( \omega \right) \right| ^2=\frac{a_1a_2(a_1+a_2)}{%
8\pi }\frac 1{\left( \omega ^2+a_1^2/4\right) \left( \omega
^2+a_2^2/4\right) } 
\end{equation}
with coefficients $a_{1,2}^2=\Gamma _0^2/2-\Omega ^2\pm \Gamma _0\sqrt{%
(\Gamma _0^2/4-\Omega ^2)}$. At a significantly weak field ($2\Omega /\Gamma
_0\ll 1$) this shape is close to the Lorentzian contour, the width $\Gamma $
of that is a quadratic function of the Rabi frequency $\Omega $: 
\begin{equation}
\label{eq2}\Gamma \approx 2\Omega ^2/\Gamma _0. 
\end{equation}
That implies a single exponential decay with time for $C(t)$. At a stronger
field ($2\Omega /\Gamma _0>1$) the shape of $\left| C\left( \omega \right)
\right| ^2$ is given by 
\begin{equation}
\label{eq3}\left| C\left( \omega \right) \right| ^2=\frac{d_2(d_2^2+d_1^2)}%
\pi \frac 1{(4\omega ^2+d_2^2-d_1^2)^2+4d_2^2d_1^2}, 
\end{equation}
where $d_1^2=\Omega ^2-\Gamma _0^2/4$ and $d_2=\Gamma _0/2$. Then the
correlation $C(t)$ shows a decaying oscillation. In the limit of a strong
field ($2\Omega /\Gamma _0\gg 1$) the shape width $\Gamma $ is invariant to
the Rabi frequency and approximates the magnitude $\Gamma _0/2$.

\section{Numerical simulation}

The properties of LSD are analized for a wide range of parameters $N$, $v$, $%
\Delta $ and $\Omega $ ($3<N<1000$, $10^{-3}<v/\Delta <10^{2}$ and $%
10^{-2}<\Omega /\Delta <10^{2}$). The total number $M$ of time pieces
covering the period interval $T_{f}$ is 32. The shape and the IPR of LSD are
determined from averaging over disorder (that is, over many random
matrices). The number of disorder realization is more than 200.

\subsection{Shape of LSD}

We restrict the study by the case of moderate strengths of state-state
interaction when the width $\Gamma $ is essentially small compared to the
whole quasienergy scale ($\Gamma \ll N\Delta /\hbar $). It implies that the
perturbation-induced variations in level density are negligible and the
quasinenergies $\omega _n$ are homogeneously distributed in the energy band $%
[-K\Delta _\omega ,K\Delta _\omega ]$ with the mean quasienergy spacing $%
\Delta _\omega =\Delta /\hbar $. Then the shape of LSD can be defined as 
\begin{equation}
\label{ShapeLSD}\rho \left( \omega \right) =\frac 1{\Delta _\omega }\frac{%
\left\langle \sum_jW_{jg}\delta (\omega -\omega _j)\right\rangle }{%
\left\langle \sum_j\delta (\omega -\omega _j)\right\rangle }, 
\end{equation}
where $\left\langle \ldots \right\rangle $ means the averaging over
disorder. We associate the shape $\rho \left( \omega \right) $ with contours
represented in equations (\ref{eq1}) and (\ref{eq3}) and determine the
coefficients $a_{1,2}$ and $d_{1,2}$ by mean square fitting of equation (\ref
{eq1}) or (\ref{eq3}) to an averaged LSD. The shape width $\Gamma $ is
estimated from the obtained coefficients as a twice halfwidth for the right
slope of $\rho \left( \omega \right) $ in the positive part of quasienergy
scale.

The numerical simulation for different magnitudes of $\Delta _\omega $, $%
\Gamma _0$, $\xi _0$ and $\Omega $ shows that the properties of $\rho \left(
\omega \right) $ are determined by some factors. First of all, that is the
mean quasienergy spacing $\Delta _\omega $. Another important factor is the
width $\Gamma _0$ of shape $\rho _0\left( \omega \right) $ for a probability
distribution of $\left| 0\right\rangle $ over quasienergies in the absence
of field. We should distinguish two localization regimes to be realized for
the distribution. A perturbative localization regime takes place when the
state-state interaction is weak and the state $\left| 0\right\rangle $ is
concentrated mainly in an individual quasienergy. For this regime the
distribution width $\Gamma _0$ is small in comparison with $\Delta _\omega $%
. On the other hand, the strong state-state coupling results in a broad
spreading of $\left| 0\right\rangle $ over the quasienergy scale. Such a
regime (designated here as a non-perturbative one) is characterized by large
magnitudes of $\Gamma _0$: $\Gamma _0>\Delta _\omega $. Certainly, the third
significant factor charachterizing the features of $\rho \left( \omega
\right) $ is the strength $\Omega $ of field-system interaction. Our study
demonstrates that the ergodic properties to be specified in terms of IPR $%
\xi _0$ for the quasienergetic distribution of $\left| 0\right\rangle $ have
no effect on the shape $\rho \left( \omega \right) $. Therefore, we consider
the shape width $\Gamma $ as a function of $\Omega $, $\Gamma _0$ and $%
\Delta _\omega $.

\begin{figure} [h!]
\epsfxsize= 8 cm
\centerline{\epsfbox{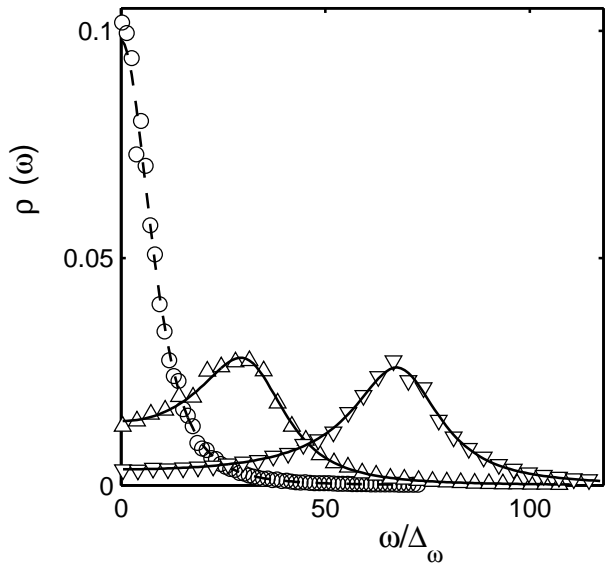}}
\vspace{2 mm}
\caption{
Shape $\rho\left( \omega \right) $ of LSD at $2\Omega
/\Gamma _{0}=0.5$ ($\bigcirc $), $2$($\Delta $), $4$($\nabla $). 
The lines show
shapes obtained from fitting of contours (\ref{eq1}) (dashed line) 
and (\ref{eq3}) (solid lines) to $\rho\left( \omega \right) $.
}
\label{Fig2}
\end{figure}

\begin{figure} [h!]
\epsfxsize= 8.5 cm
\centerline{\epsfbox{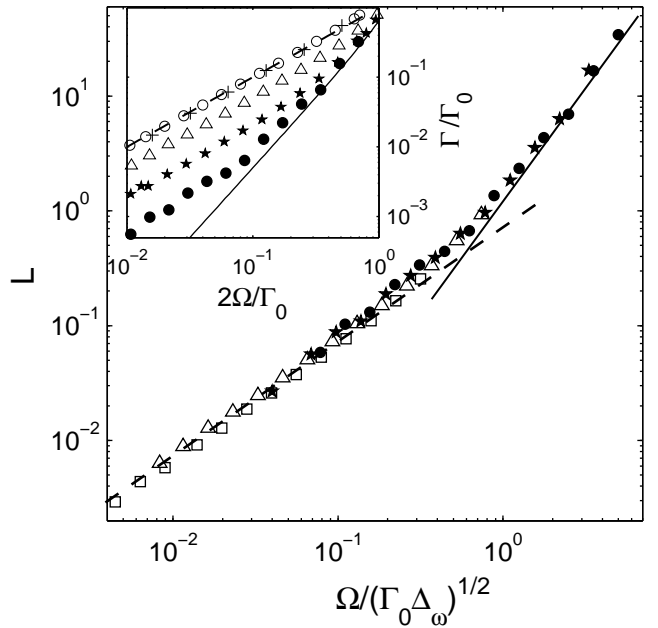}}
\vspace{2 mm}
\caption{
The localization length $L$ as a function of 
$\Omega /\sqrt{\Delta _{\omega }\Gamma _{0}}$ at a weak field 
($2\Omega /\Gamma _{0}<1$) for $\Gamma_{0}/\Delta _{\omega }=0.9$ 
(squares), 2.5 ($\nabla$), 20 (stars)
and 225 (full circles). The dashed line gives approximation (\ref{eq5}) obtained for
the points with $\Gamma _{0}/\Delta _{\omega }>1$. The solid line represents 
law (\ref{eq2}) predicted by the TSSIL model. The inset shows the ratio $\Gamma
/\Gamma _{0}$ as a function of $2\Omega /\Gamma _{0}$ at $2\Omega
/\Gamma _{0}<1$ for $\Gamma _{0}/\Delta _{\omega }=0.06$ ($+$), 
$0.2$ ($\bigcirc $), $2.5$ ($\Delta $), 20. (stars) and 226 (full circles).
The dashed line corresponds to dependence (\ref{eq4}). The solid line represents 
law (\ref{eq2}) predicted by the TSSIL model.
}
\label{Fig3}
\end{figure}

Similar to the TSSIL model, at a weak field ($2\Omega /\Gamma _0<1$) the
shape $\rho \left( \omega \right) $ is shown in Figure \ref{Fig2} to be well
described by contour (\ref{eq1}). The obtained values of $\Gamma $ satisfy
the requirement $\Omega ^2/\Gamma _0\leq \Gamma \leq 2\Omega $. The type of $%
\Omega $-dependence for the width $\Gamma $ is specified by relation between 
$\Delta _\omega $, $\Gamma _0$ and $\Omega $. Our study demonstrates that
the low values of $\Gamma _0$ or $\Omega $ give the linear low for the
dependence. At $\Gamma _0<\Delta _\omega $ the width $\Gamma $ is seen from
the inset of Figure \ref{Fig3} to be a linear $\Gamma _0$-independent
function of $\Omega $: 
\begin{equation}
\label{eq4}\Gamma \approx 2\Omega .
\end{equation}
This dependence can be considered as a manifestation of the perturbative
regime for distribution $\rho _0\left( \omega \right) $. At $\Gamma
_0>\Delta _\omega $ (when a non-perturbative localization regime for $\rho
_0\left( \omega \right) $ is realized) the increase of $\Omega $ results in
the transition of $\Omega $-dependence for $\Gamma $ from the linear law to
the quadratic one. We associate such a behavior of the width $\Gamma $ with
the magnitude of localization length $L$ for LSD. Figure \ref{Fig3} shows
that a linear function of $\Omega $%
\begin{equation}
\label{eq5}\Gamma \approx A\Omega \sqrt{\Delta _\omega /\Gamma _0}
\end{equation}
takes place when $L\ll 1$. The coefficient $A=0.7$ is obtained from mean
square fitting of (\ref{eq5}) to calculated values of $\Gamma $. On the
other hand, at $L\gg 1$ the $\Omega $-dependence for the quantity $\Gamma $
is close to the quadratic low (\ref{eq2}) predicted by the TSSIL model. As
for distribution $\rho _0\left( \omega \right) $, the transition from the
linear to quadratic $\Omega $-dependence of $\Gamma $ can be associated with
crossover of the shape $\rho \left( \omega \right) $ from the perturbative
to non-perturbative localization regime.

\begin{figure} [h!]
\epsfxsize= 8.5 cm
\centerline{\epsfbox{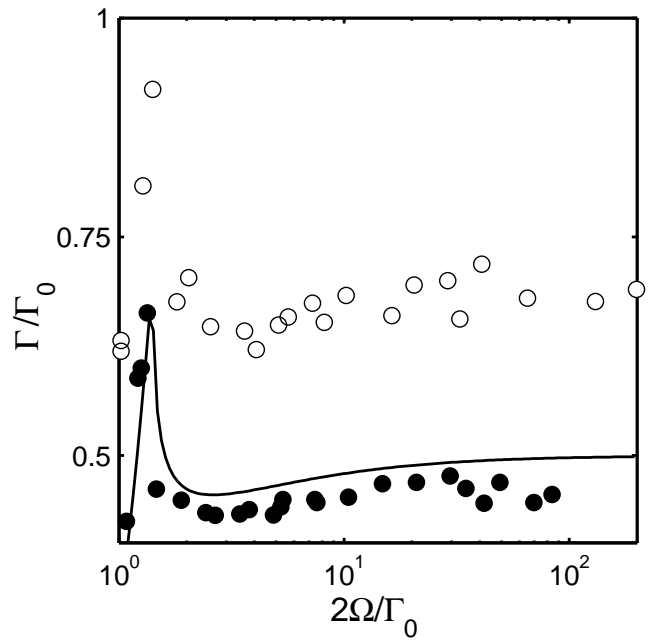}}
\vspace{3 mm}
\caption{
The ratio $\Gamma/\Gamma _{0}$ as a function of $2\Omega
/\Gamma _{0}$ at a strong field ($2\Omega /\Gamma _{0}>1$) 
for $\Gamma _{0}/\Delta _{\omega
}>1$ (full circles) and $\Gamma _{0}/\Delta _{\omega }<1$ ($\bigcirc $).
The solid line shows a relevant dependence predicted by the TSSIL model.
}
\label{Fig4}
\end{figure}

At a strong field ($2\Omega /\Gamma _0>1$) the shape $\rho \left( \omega
\right) $ is shown in Figure \ref{Fig2} to be well fitted by contour (\ref
{eq3}) obtained from the TSSIL model. As for the case of weak field, the
behavior of $\Omega $-dependence for the shape width $\Gamma $ is determined
by the localization regime realized for shape $\rho _0\left( \omega \right) $%
. Figure \ref{Fig4} demonstrates that at $\Gamma _0>\Delta _\omega $ the $%
\Omega $-dependence of $\Gamma $ is in good agreement with a relevant curve
obtained from the TSSIL model. At high field amplitudes (when $2\Omega
/\Gamma _0\gg 1$) the width $\Gamma $ approximates the value $\Gamma _0/2$.
The minor discrepancy between $\Gamma $ and the curve predicted by the TSSIL
model can be explained by a divergence of the temporal profile of field from
the monochromatic shape. At the perturbative regime of $\rho _0\left( \omega
\right) $ the $\Omega $-dependence of $\Gamma $ is proportional to the TSSIL
curve. Our analysis shows that for any magnitude of $\Omega $ the calculated
values of $\Gamma $ are higher then ones from the TSSIL model by a factor of
1.4. In the limit of high field amplitudes the width $\Gamma $ is close to
the magnitude $\Gamma _0/1.4$.

\subsection{Inverse participation ratio of LSD}

\begin{figure} [h!]
\epsfxsize= 8.5 cm
\centerline{\epsfbox{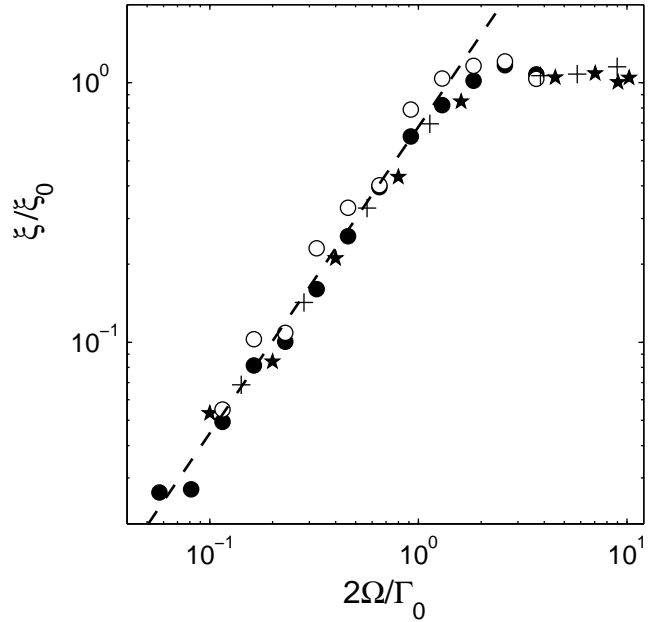}}
\vspace{3 mm}
\caption{
Dependence of the IPR $\xi$ on the ratio $2\Omega /\Gamma _{0}$ 
at $\Gamma _{0}/\Delta _{\omega }=40$ ($\xi _{0}=30$ ($+$), $40$(stars))
and $\Gamma _{0}/\Delta _{\omega }=100$ ($\xi _{0}=39$ ($\bigcirc $), $100$
(full circles)). The dashed line shows approximation (\ref{eq6}). 
}
\label{Fig5}
\end{figure}

We investigate the IPR $\xi $ under the circumstances of localized regime
for the probability distribution of basis state $\left| 0\right\rangle $
over the eigenstates of ${\bf H}^{(0)}$. It implies that the state $\left|
0\right\rangle $ is spread over a sufficiently large number of the
eigenstates and $N\gg \xi _0\gg 1$. Here the quantity $\xi _0$ is the IPR
for the probability distribution of $\left| 0\right\rangle $ and gives the
actual number of eigenstates involving the state $\left| 0\right\rangle $ 
\cite{ref21}. In the study the main attention is paid to the case of $\xi
\gg 1$. We accept the IPR as $\xi =(\left\langle
\sum_j|W_{jg}|^2\right\rangle )^{-1}$ and obtain this quantity from
numerical simulation. Here $\left\langle \ldots \right\rangle $ means the
averaging over disorder. The IPR $\xi $ is considered as a function of $%
\Omega $, $\xi _0$ and $\Gamma _0$. Results of the study are represented in
Figure \ref{Fig5}.

Our study demonstrates that at fixed $\Omega $ and $\Gamma _0$ the IPR $\xi $
varies proportionally with the parameter $\xi _0$. This fact testifies that
the change in $\xi $ with the Rabi frequency is due to transformations in
the shape of LSD. For instance, in the limit of strong field ($2\Omega
/\Gamma _0\gg 1$) the width $\Gamma $ of the shape $\rho \left( \omega
\right) $ is $\Omega $-independent and $\Gamma \approx \Gamma _0/2$. As a
result, the quantity $\xi $ is also invariant to $\Omega $ and approximates
the magnitude of $\xi _0$ (see Figure \ref{Fig5}). In other words, the IPR
of LSD is equal to the IPR for the distribution of $\left| 0\right\rangle $
over the eigenstates of ${\bf H}^{(0)}$ (here we should take into account
the presence of two peaks of LSD in the negative and positive parts of
quasienergy spectrum). As indicated earlier, at a weaker field ($2\Omega
/\Gamma _0\lesssim 1$) the variation in $\Omega $ give rise to essential
shape transformations both in the shape width and in the contour form. The
analysis of calculated data shows that the behavior of IPR for this
parameter region can be empirically approximated by the law: 
\begin{equation}
\label{eq6}\xi \approx B\xi _0(\Omega /\Gamma _0)^\beta ,
\end{equation}
where the coefficients $B=1.54$ and $\beta =1.18$ are obtained from mean
square fitting of (\ref{eq6}) to calculated values of $\xi $. 

Notice that the 
observed $\Omega$-dependence of IPR is close to the linear law and differs 
significantly from the expected quadratic dependence to be realized 
in the case of the Lorentzian shape for LSD. Probably, this difference is
attributed to the fact that we are not far enough in the asymptotic regime 
of weak field implying $2\Omega/\Gamma_{0}\ll1$. In our simulation 
we have $2\Omega/\Gamma_{0}\gtrsim0.1$ and the shape of LSD 
appers to can not be well reduced to the Lorentzian contour. 
Unfortunately, the numerical simulation of localized regime for LSD 
with $2\Omega/\Gamma_{0}<0.1$ requires too large matrix sizes 
and much computational efforts being beyond our numerical abilities.

\section{Conclusion}

We have analyzed localization and ergodic properties of LSD for a
periodically excited generic system of coupled quantum states with strong
imperfection in the unperturbed energies. These properties have been
demonstrated to be determined essentially by a state-state interaction
resulting in the localization effect for the system in the absence of
external field. If the interaction is so strong that the conservative system
exhibits this effect the shape of LSD can be obtained from a relevant model
of driven two-state system with irreversible losses. In a sense the
state-state interaction may be recognized as the losses acting till the
mesoscopic effect becomes evident. The time scale and the rate of a
correlation decay caused by such losses are specified in terms of the shape
width and the IPR of LSD.

Author thanks V. Churakov for useful discussions and remarks.

\end{document}